\documentclass{PoS}
\usepackage{amsmath}
\usepackage{mathrsfs}
\usepackage{mathtools}
\usepackage{slashed}

\newcommand{\fsl}[1]{\ensuremath{\mathrlap{\!\not{\phantom{#1}}}#1}}
\newcommand{\tmop}[1]{\ensuremath{\operatorname{#1}}}

\title{$K_L$ - $K_S$ mass difference computed with a 171 MeV pion mass}

\ShortTitle{$K_L$ - $K_S$ mass difference computed with a 171 MeV pion mass}

\author{\speaker{Ziyuan Bai} \\
	Department of Physics, Columbia University, New York\\
E-mail: \email{zb2174@columbia.edu}}

\abstract{
	In this work, I used a $32^3 \times 64 \times 32$, 2+1 flavor domain wall
	lattice with Iwasaki+DSDR gauge action. The pion mass is 171 MeV and the
	kaon mass is 492 MeV. We implement the Glashow-Iliopoulos-Maiani (GIM)
	cancellation using charm quark masses of 750 MeV and 592 MeV. This is an
	intermediate calculation, in that we are using both a coarse lattice spacing
	(1/a = 1.37GeV) so we expect significant discretization error coming from
	charm quark mass and we are also using unphysical kinematics for the pion.
	The main purpose of this calculation is to study the contribution from
	the two-pion intermediate state when the energy of a two-pion state is
	lower than that of the kaon, as well as the corresponding finite volume correction 
to the $\Delta M_K$.}

\FullConference{The 32nd International Symposium on Lattice Field Theory\\
	23-28 June, 2014\\
Columbia University New York, NY}

\begin{document}

\section{Introduction}
The $K_L - K_S$ mass difference $\Delta M_K$, with an experimental value of
$3.483(6)\times10^{−12}$ MeV is an important quantity in particle physics, for
the reason that it leads to the prediction of charm quark mass scale, and it's 
small size provides an important test for the Standard Model. Perturbation theory calculation
fails to make a convincing prediction of $\Delta M_K$. As pointed out in 
\cite{NNLO_klks}, the size of NNLO is about 0.36 of the size of LO contribution, 
and the purely non-perturbative, long distance part, is estimated
to be about 30\% of the total contribution. Therefore, lattice QCD is the only 
reliable way to calculate $\Delta M_K$ in the Standard Model, with all systematic 
error controlled. 

Previous attempts to calculate $\Delta M_K$\cite{Yu:klks16}\cite{Yu:klks24} are encouraging.
These are done using an unphysically large $329$ MeV pion mass. If we go to physical or near physical pion mass,
two issues might arise: the treatment of the two-pion intermediate state which gives
an exponential growing contribution to our integrated correlator, as well as the corresponding finite
volume correction. This work is done primarily to
address these issues.  This is an	intermediate calculation, in that we are using both a 
coarse lattice spacing(1/a = 1.37GeV) so we expect significant discretization error 
coming from	the charm quark mass and that we are also using unphysical kinematics for the pion.

\section{Evaluation of $\Delta M_K$ on the Lattice}
\begin{figure}[h]
	\centering
	\includegraphics[width=0.6\textwidth]{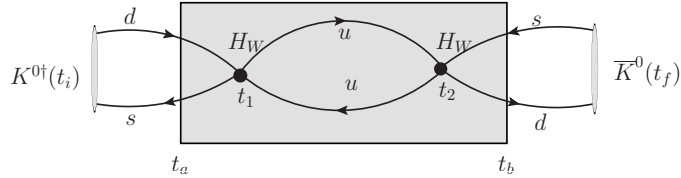}
	\caption{Example diagram for $K^0 - \bar{K^0}$ mixing. The two weak Hamiltonians are 
	integrated over the time interval $[t_a, t_b]$.}
	\label{fig:k_mix}
\end{figure}

The $K^0 - \bar{K^0}$ mixing is represented in Figure \ref{fig:k_mix}. We have 
the $\Delta S = 1$ weak Hamiltonian $H_W$:
\begin{eqnarray}
	H_W  &=& \frac{G_F}{2} \sum\limits_{q,q' = u,c} V_{qd} V^{*}_{q's} 
	( C_1 Q_1^{qq'} + C_2 Q_2^{qq'} )\\
	Q_1^{qq'} &=& (\bar{s}_i d_i )_{V-A} (\bar{q}_j q'_j )_{V-A} \\
	Q_2^{qq'} &=& (\bar{s}_i d_j )_{V-A} (\bar{q}_j q'_i)_{V-A}\,. 
\end{eqnarray}

By integrating the time-ordered product of the two $\Delta S = 1$ operator, 
we can obtain the integrated correlator for this $\Delta S= 2$ process.
The integrated correlator is defined as:
\begin{equation}
	\mathcal{A} = \frac{1}{2} \sum\limits_{t_2 = t_a}^{t_b} \sum\limits_{t_1 
	= t_a}^{t_b} \langle 0 | T \left\{ \overline{K}^0(t_f) 
	H_W(t_2) H_W(t_1) \overline{K}^0 (t_i) \right\} | 0 \rangle \,.
\end{equation}

If we insert a complete set of intermediate states, we can find: 
\begin{equation}
	\mathcal A = N_{K}^{2} e^{-M_{K} ( t_{f} -t_{i} )} \left\{ \sum_{n }
		\frac{\langle \bar{K}^{0} | H_{w} | n \rangle \langle n | H_{w} | K^{0}
		\rangle}{M_{K} -M_{n}} \left( -T+ \frac{e^{( M_{K} -M_{n} ) T}-1}{M_{K}
	-M_{n}} \right) \right\}\,.
\end{equation}

The term proportional to $T$ can be related to $\Delta M_K$:
\begin{equation}
	\Delta M_k = 2 \sum \limits_{n } \frac{\langle \bar{K}^{0} | H_{w} | 
	n \rangle \langle n | H_{w} | K^{0}	\rangle}{M_{K} -M_{n}} \,.
		\label{Eq.ind_contr}
\end{equation}

In order to extract the linear term with $T$ from the integrated correlator $\mathcal A$, we have
to deal with the second term which involves an exponential contribution. Our choice for $T$
is enough to make most of the intermediate state with energy higher than the kaon highly suppressed 
(except for the $\eta$). For intermediate states that have lower energy than the kaon, we must 
subtract their exponentially growing contribution to our integrated correlator. We have two ways of doing this:
we can either directly calculate the matrix element and determine their exponential contribution:
\begin{equation}
	N_{K}^{2} e^{-M_{K} ( t_{f} -t_{i} )} 
		\frac{\langle \bar{K}^{0} | H_{w} | n \rangle \langle n | H_{w} | K^{0}
		\rangle}{(M_{K} -M_{n})^2} e^{( M_{K} -M_{n} ) T}\, ,
\end{equation}
and subtracted it. Or, we can add a scalar operator $\bar{s}d$, or a pseudoscalar 
operator $\bar{s}\gamma_5 d$, to our weak Hamiltonian, chosen to make their contribution 
disappear. We have this freedom because these two operators can be written as a total
divergence of a vector or axial current, and therefore they will not affect the physical linear term in 
our integrated correlator. 

In this calculation, the states lighter than the kaon are single pion state and two-pion
state with either isospin $0$ or $2$. The $\eta$ meson is slightly heavier than the kaon, 
but it's energy difference is not enough to make it highly suppressed, so we also have 
to subtract the $\eta$ state. Because we have disconnected diagrams, we must also subtract 
vacuum state, which might be a major source of statistical noise.

\section{Details of simulation}
We work on a $2+1$ flavor, $32^3 \times 64 \times 32$ DWF lattice, with the Iwasaki + DSDR
gauge action, and an inverse lattice spacing $1/a = 1.37$ GeV. 
The pion mass is $171$ MeV and the kaon is $492$ MeV.
We implement GIM cancellation by including a quenched charm quark. We use
two choices of charm quark mass, $0.38$ and $0.3$ in lattice units, which correspond to $750$ MeV and $592$ MeV.
One might think that $0.38$ is too high because it can produce an unphysical state 
that propagates on the $5$th dimension. However, because we are only interested in the physics 
on the domain wall of $5$th dimension, which couples weakly to this unphysical state, having
a charm mass of 0.38 will not give rise to much systematic error. In order to accelerate the inversion,
we used low-mode deflation with 580 eigenvectors obtained using the Lanczos algorithm. Also,
we use Mobius fermions with $b+c=2.667$, $L_s = 12$, which leaves our residual mass 
unchanged from its unitary value.
We are using $405$ configurations, about twice the number presented in the talk.

\begin{figure}[h]
	\begin{center}
		\begin{tabular}{c|c}\hline
			\includegraphics[width=0.3\textwidth,height = 0.1\textwidth]{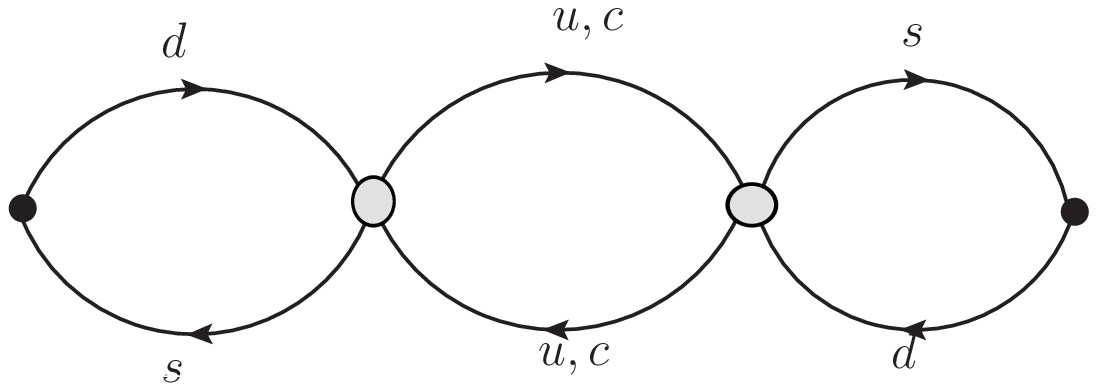} type 1&
			\includegraphics[width=0.3\textwidth,height = 0.1\textwidth]{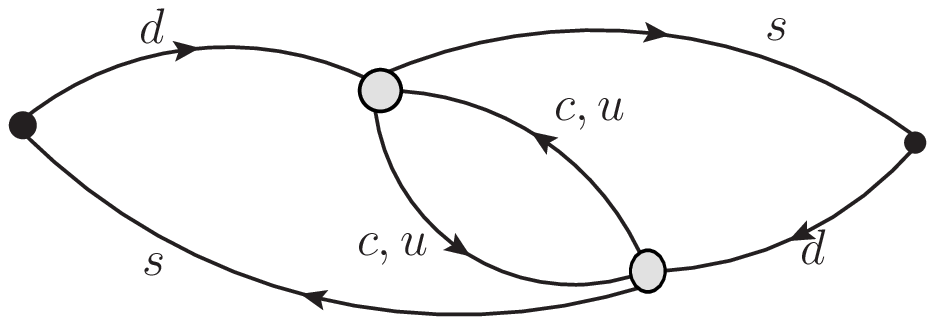} type 2\\\hline
			\includegraphics[width=0.3\textwidth,height = 0.1\textwidth]{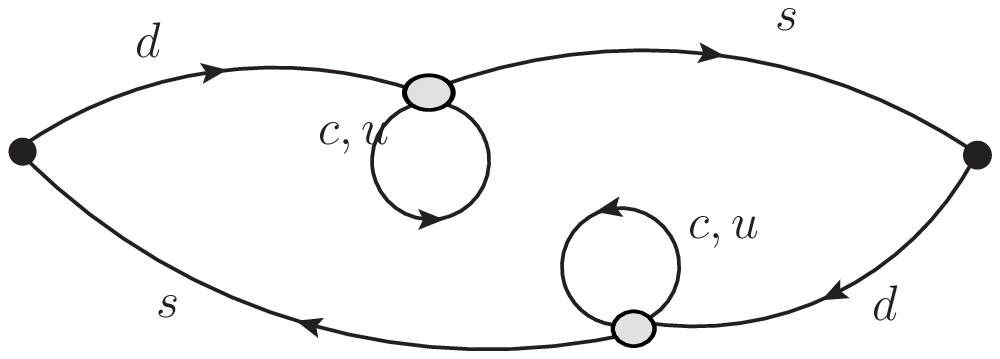} type 3 &
			\includegraphics[width=0.3\textwidth,height = 0.1\textwidth]{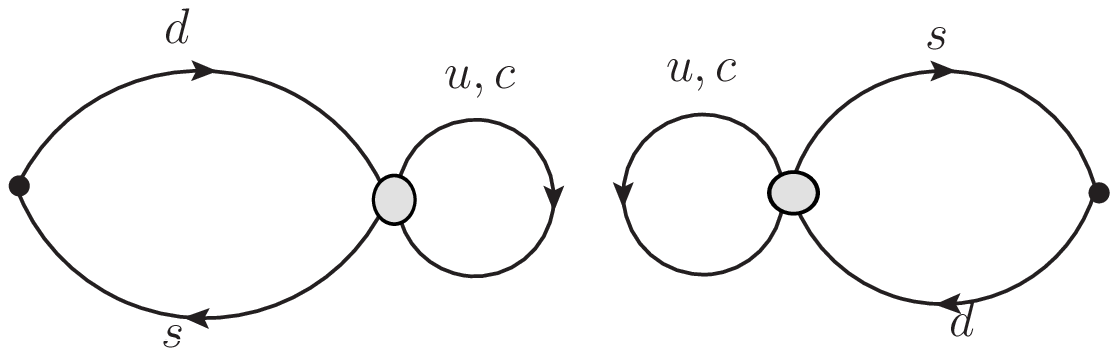} type 4\\
			\hline
		\end{tabular}
		\caption{Four types of four point diagrams in the calculation of integrated correlator}
	\label{fig:4pt}
	\end{center}
\end{figure}

\begin{figure}[h]
	\centering
	\begin{tabular}{c|c}\hline
		\includegraphics[width=0.3\textwidth,height = 0.11\textwidth]{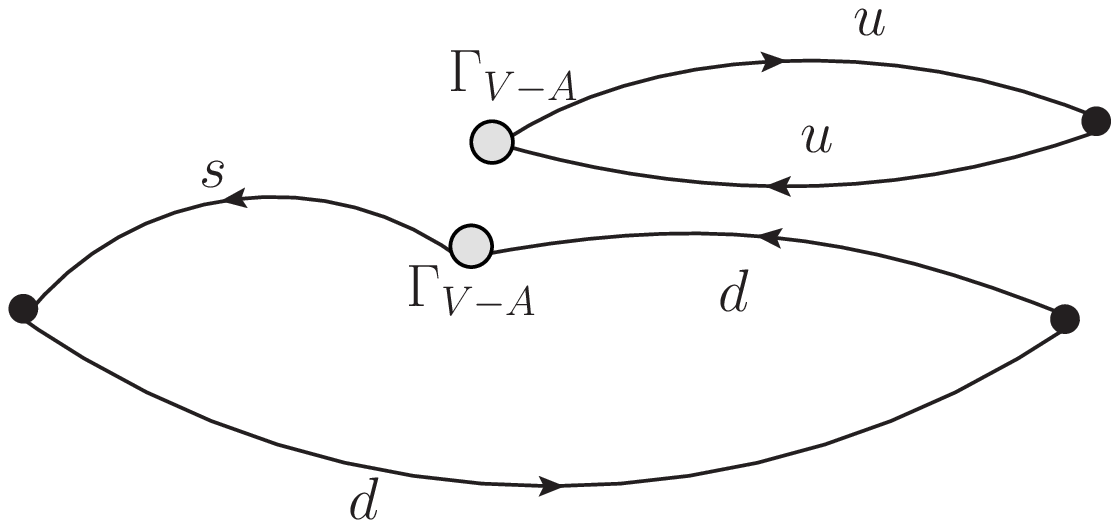} type 1&
		\includegraphics[width=0.3\textwidth,height = 0.11\textwidth]{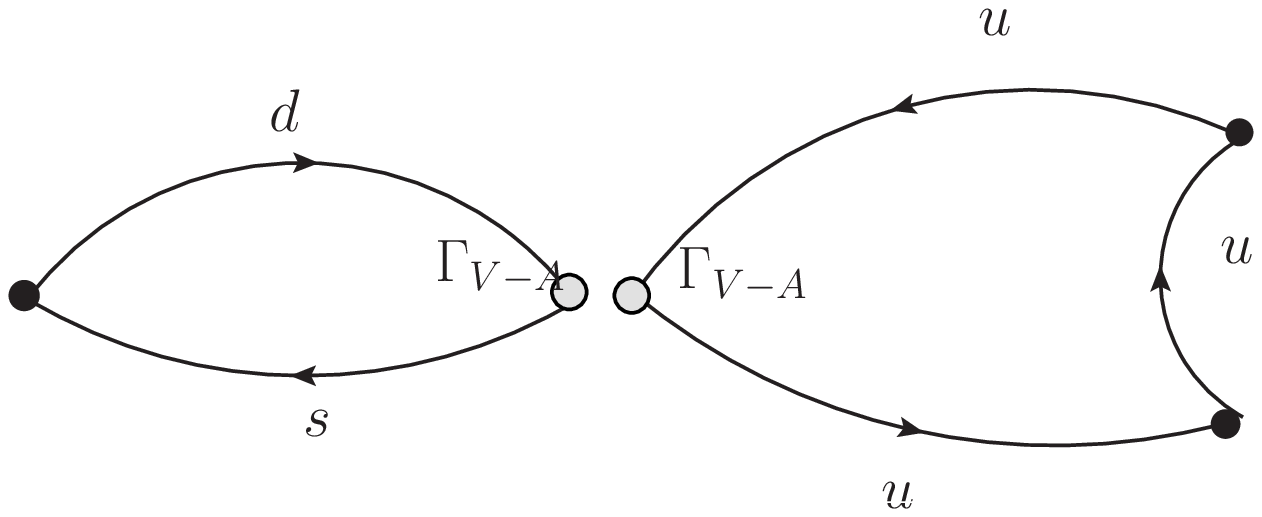} type 2\\ \hline
		\includegraphics[width=0.3\textwidth, height = 0.11\textwidth]{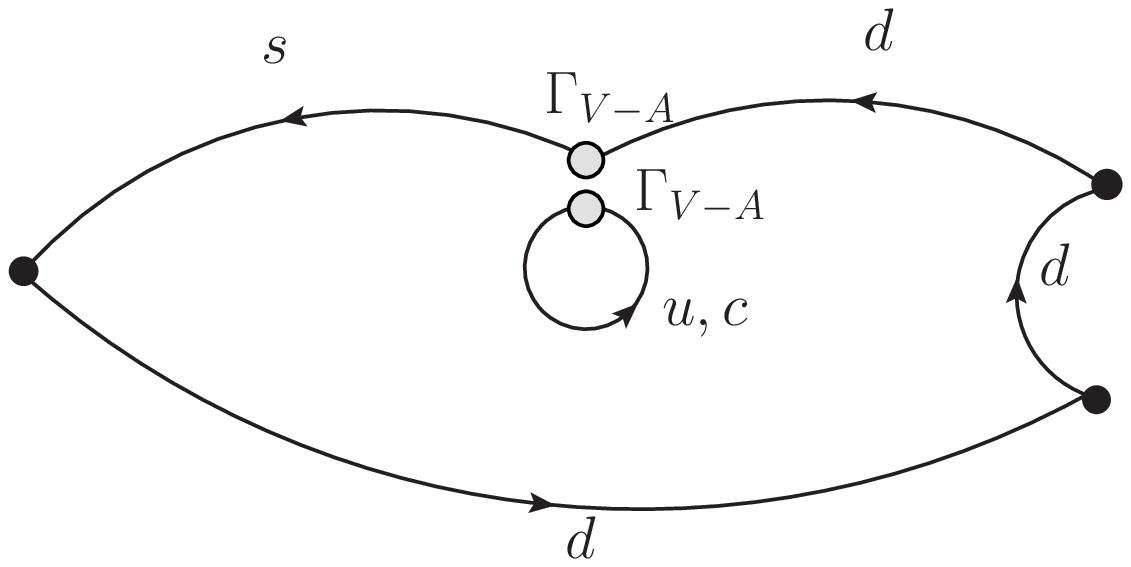} type 3 &
		\includegraphics[width=0.3\textwidth, height = 0.11\textwidth]{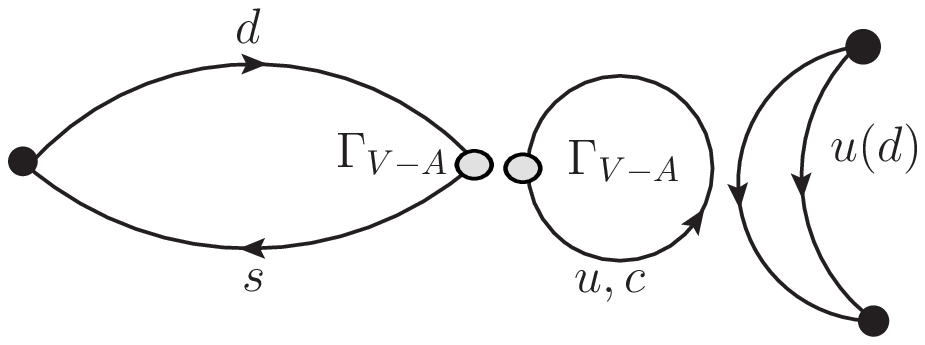} type 4\\\hline
	\end{tabular}
	\caption{Four types of diagrams in the calculation of $\langle \pi\pi | H_W | K \rangle$}
	\label{fig:k2pipi}
\end{figure}

We calculate all the four point diagrams corresponding to the $K^0 - \bar{K^0}$ mixing
process, as shown in Figure \ref{fig:4pt}. In order to subtract the two-pion intermediate state,
we must also calculate the kaon to two-pion matrix element $\langle\pi\pi| H_W|K\rangle$, 
as shown in Figure \ref{fig:k2pipi}.
We use Coulomb gauge fixed wall source for the kaon, and a point source propagator
at each time slice for the internal quark lines coming from one of the weak vertex 
in type 1/2 four point diagrams. 
For the self-loop in type 3/4 four point diagrams and type 3/4 kaon to two-pion diagrams, 
we use a random space-time volume source with $80$ hits. This can significantly 
reduce the number of inversions required compared to the random wall source with 5 hits that we used
in \cite{Yu:klks24}, while leaving the statistical error the same size.
To suppress vacuum noise in the computation of $\langle\pi\pi| H_W|K\rangle$, 
we separate the two-pion in the sink by $4$ units in time direction. 

\section{Results and finite volume correction}
Before we do the linear fitting to obtain $\Delta M_K$, we have to first subtract
all the intermediate states with energy lower than or close to the kaon mass. Firstly, because the vacuum 
intermediate state has the largest matrix element $\langle 0 | H_W |K \rangle$, if we
directly subtract the vacuum state, then the statistical error is too large in our 
final integrated correlator, making fitting impossible. Therefore, we must use our freedom
of adding a pseudoscalar operator $\bar{s}\gamma_5d$ to eliminate the vacuum contribution. 
Because $\langle \eta| H_W |K \rangle$ has the largest statistical error coming from the disconnected
diagrams, we use a scalar operator $\bar{s} d$ to eliminate the $\eta$ contribution. We 
tune the coefficient $c_s$ and $c_p$ in our modified Hamiltonian $H_W' = H_W + c_p \bar{s}
\gamma_5 d + c_s \bar{s} d$, so that 
\begin{eqnarray}
	\langle 0 | H_W + c_p \bar{s}\gamma_5 d| K\rangle &=& 0 \,, \\
	\langle \eta | H_W + c_s \bar{s} d| K\rangle &=& 0 \,.
\end{eqnarray}
\begin{table}
	\centering
	\begin{tabular}{c|c|c|c|c} \hline
		$m_c$ & $c_{1s}$ & $c_{2s}$ & $c_{1p}$ &$c_{2p}$ \\ \hline
		750 MeV & $6.4(14)\times10^{-4}$ & $-8.2(10)\times10^{-4}$
		 & $-4.356(14)\times10^{-4}$ & $7.567(15)\times10^{-4}$ \\ \hline
		592 MeV & $5.7(15)\times10^{-4}$ & $-7.2(12)\times10^{-4}$
		 & $-4.052(18)\times10^{-4}$ & $6.626(18)\times10^{-4}$ \\ \hline
	\end{tabular}
	\caption {The subtraction coefficient $c_s$, $c_p$, for $Q_1$ and $Q_2$ separately. $m_c=0.38,0.30$
	respectively on lattice}
	\label{table:subtr}
\end{table}
The subtraction coefficients $c_s$, $c_p$ are shown in Table \ref{table:subtr}.
With the modified Hamiltonian, we calculate the kaon to two-pion matrix element
$\langle \pi\pi| H_W' | K \rangle$, with both isospin $0$ and $2$, the results are given
in Table \ref{table:k2pipi}.

\begin{table}[h]
	\centering
	\begin{tabular}{c|c|c|c|c}\hline
	$m_c$ & $\langle \pi\pi_{I = 2} | Q_1 | K \rangle$ & $\langle \pi\pi_{I = 2} | Q_2 | K \rangle$
		& $\langle \pi\pi_{I = 0} | Q_1 | K \rangle$ & $\langle \pi\pi_{I = 0} | Q_2 | K \rangle$ \\\hline
		$750$ MeV & $1.291(8)\times 10 ^{-4}$ & $1.291(8)\times 10 ^{-4}$
		& $-5.2(35)\times 10 ^{-4}$ & $7.2(31)\times 10 ^{-4}$ \\\hline
		$592$ MeV & $1.284(10)\times 10^{-4}$ & $1.284(10)\times 10^{-4}$ &
	$-4.4(34)\times10^{-4}$ & $8.7(28) \times 10^{-4}$ \\ \hline
	\end{tabular}
	\caption{Kaon to two-pion matrix elements. The fact that the two $I=2$ matrix element
	are exactly the same is not surprising because they result from the same diagrams.}
	\label{table:k2pipi}
\end{table}

\begin{figure}[h]
	\begin{tabular}{cc}
			\includegraphics[width=0.5\textwidth]{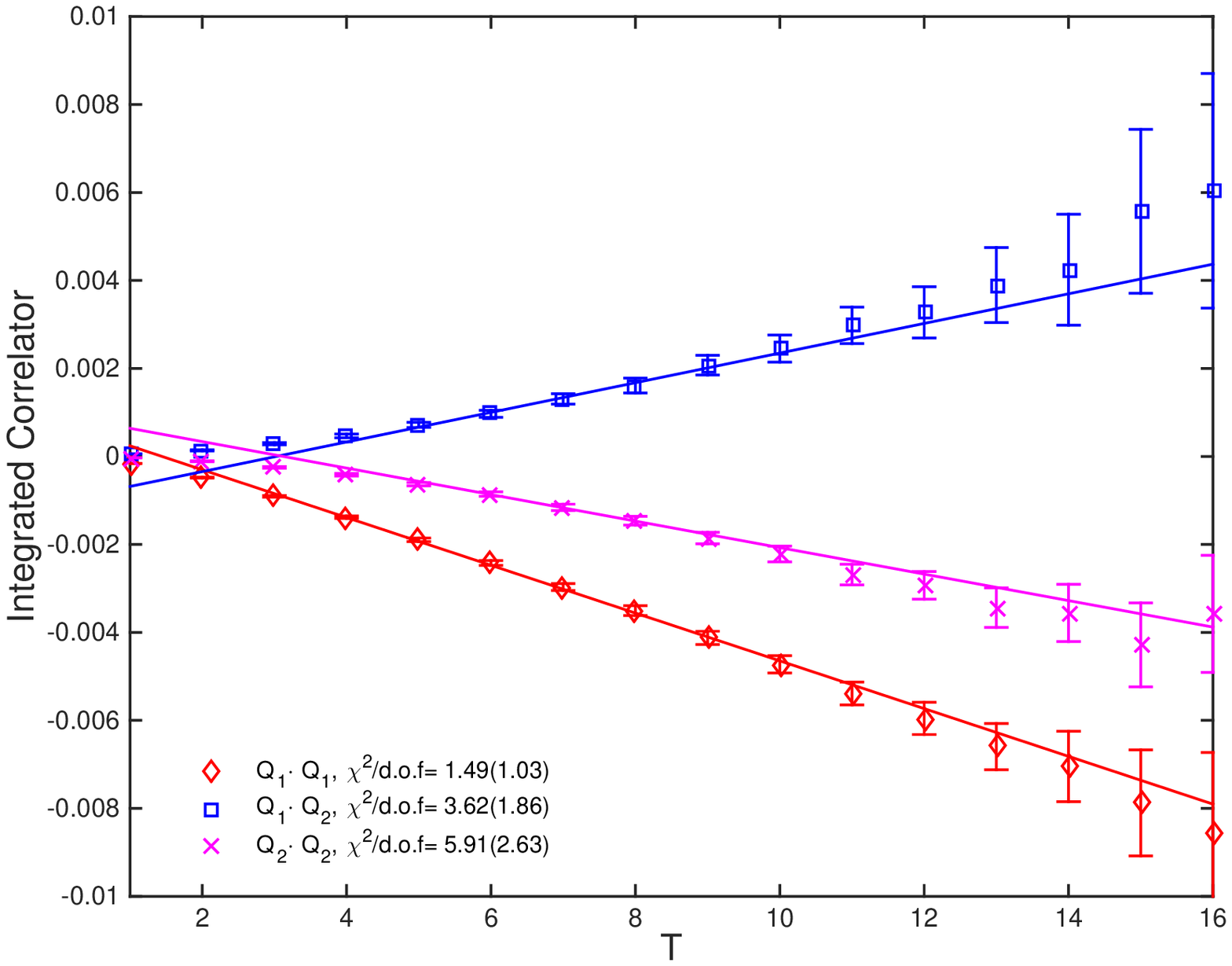} &
			\includegraphics[width=0.5\textwidth]{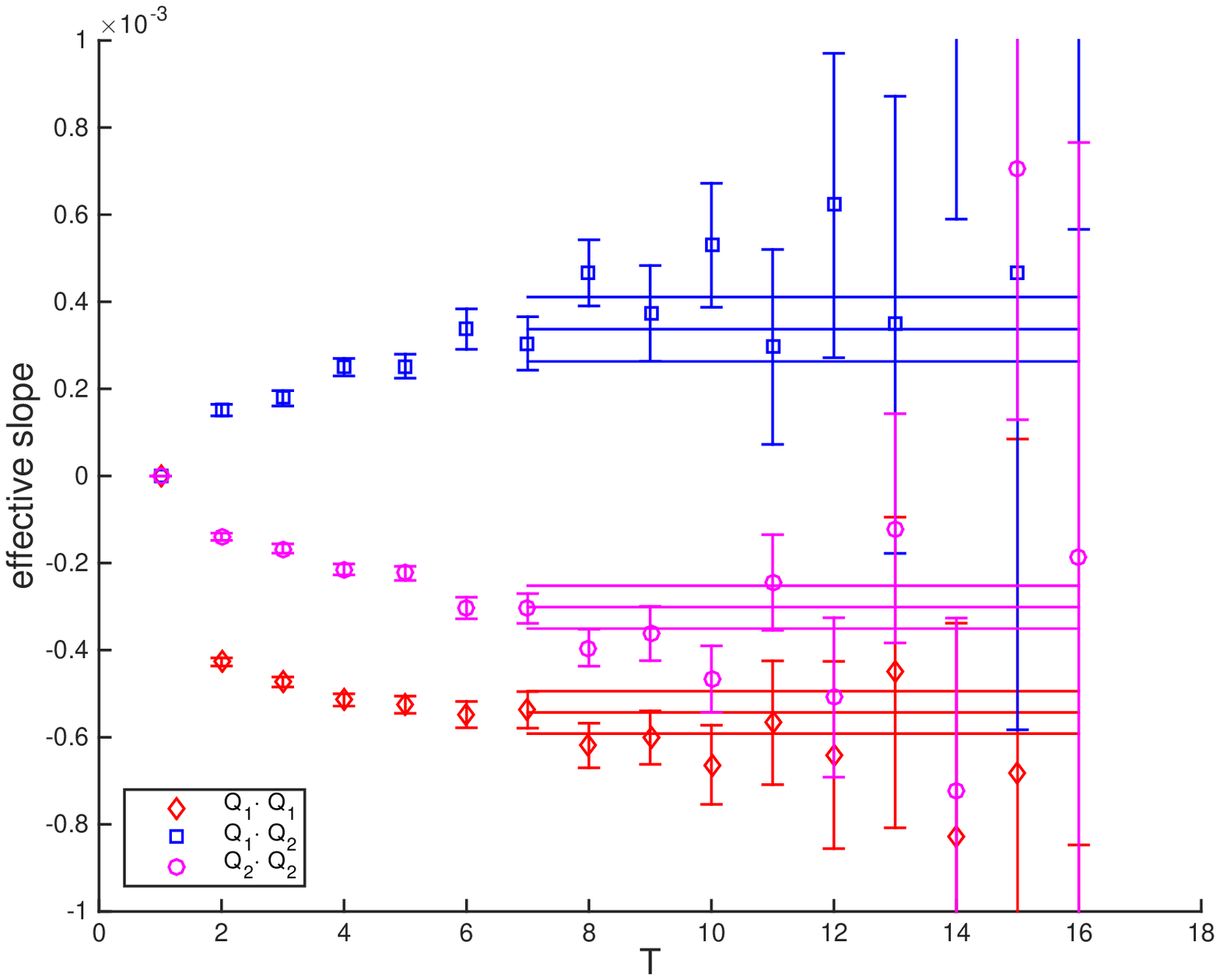}\\
	\end{tabular}
	\caption{Integrated correlator using $m_c=0.38$, starting fitting range 
		$T_{min} = 7$, and the corresponding effective slope plot.}
	\label{fig:intcorr}
\end{figure}

After we have subtracted all the exponentially growing intermediate state contributions,
we can fit our integrated correlator 
as a linear function of $T$. This is shown in Figure \ref{fig:intcorr}, as well as the 
corresponding effective slope. We show the individual contributions to $\Delta M_K$,
in Table \ref{table:ind_klks}. All of these numbers have been multiplied by the 
Wilson coefficient at the energy scale $3$ GeV, which can be found in Table \ref{table:WC}. 
\begin{table}[h]
	\centering
	\begin{tabular}{c|c|c|c|c|c}\hline
		$m_c$ & $T_{min}$ & $Q_1 Q_1$ & $Q_1 Q_2$ & $Q_1 Q_2$ & $\Delta M_K$\\ \hline
		$0.30$ & $6$ & $ 0.59(5) $ & $1.32(17)$ & $2.58(36)$ & $4.50(49)$ \\ \hline 
		$0.30$ & $7$ & $ 0.56(6) $ & $1.20(23)$ & $2.72(56)$ & $4.50(75)$ \\ \hline 
		$0.30$ & $8$ & $ 0.67(9) $ & $1.77(31)$ & $3.39(59)$ & $5.83(85)$ \\ \hline 
		$0.38$ & $6$ & $ 0.73(4) $ & $1.37(20)$ & $3.07(32)$ & $5.17(47)$ \\ \hline 
		$0.38$ & $7$ & $ 0.70(6) $ & $1.22(27)$ & $3.04(50)$ & $4.96(73)$ \\ \hline 
		$0.38$ & $8$ & $ 0.79(8) $ & $1.70(34)$ & $3.72(65)$ & $6.22(94)$ \\ \hline 
	\end{tabular}
	\caption{$\Delta M_K$ and the individual contribution from different operator combinations,
	in the units of $10^{-12}$ MeV. We have varied the starting point of our fitting range. 
All numbers are multiplied by the appropriate Wilson coefficient computed in the
$(\gamma_\mu, \gamma_\mu)$ scheme.}
	\label{table:ind_klks}
\end{table}

We calculated the Wilson coefficient at $3$ GeV to reduce error in the perturbative calculation, 
but it might be too high for our coarse lattice with $1/a=1.37$ GeV. Therefore, we first match 
our lattice operator non-perturbatively to a regularization independent, Rome-Southampton 
scheme \cite{Martinelli199581,PhysRevD.68.114506} with non-exceptional momentum 
under 1.5 GeV. We used both the RI/SMOM($\gamma_{\mu},\fsl{q}$) and RI/SMOM($\gamma_{\mu},
\gamma_{\mu}$) schemes \cite{PhysRevD.84.014001}. Then we perform step scaling, i.e. 
by matching to higher energy scale using a finer intermediate lattice, and finally we match 
to $3$ GeV in the RI/SMOM scheme. 
The matching factors from RI/SMOM to $\overline{MS}$ is given by $1 + \Delta r$
\footnote{C.Lehner and C. Sturm, private communications}
, and the $\Delta r$
in four-flavor theory can also be found in Table \ref{table:WC}.
The $\overline{MS}$ Wilson coefficient at $3$ GeV can be calculated 
using equations in \cite{Buras}. 
\begin{table} [h]
	\begin{center}
		\begin{tabular}{c c|c c|c c|c c}\hline
			 $C_1^{\overline{MS}}$ &$C_2^{\overline{MS}}$ & $\Delta r_{11}=\Delta r_{22}$ &
			$\Delta r_{12} = \Delta r_{21}$ & $Z_{11}=Z_{22}$ &
			$Z_{12}=Z_{21}$ &$C_1^{lat}$ & $C_2^{lat}$ \\ \hline
			 -0.2394& 1.1068& -0.0566&0.0065 & 0.5589 & -0.0918
			&-0.2179 & 0.6027 \\\hline
			-0.2394& 1.1068& -0.0022&0.0065 & 0.5008 & -0.0819 &-0.2064 & 0.5713 \\\hline
		\end{tabular}
		\caption{The $\overline{MS}$ Wilson Coefficients, $RI/SMOM
			\rightarrow \overline{MS}$ matching matrix $\Delta r$, 
			 $lat\rightarrow RI$ matching matrix $Z$ obtained using $Z_q$ calculated in different 
		schemes, and finally the lattice Wilson Coefficient at scale 3.0GeV. 
		1st row: $(\gamma_\mu, \slashed{q})$ scheme, 2nd row: $(\gamma_\mu, \gamma_\mu)$.
		We didn't include statistical error because all less than 1\%.}
		\label{table:WC}
	\end{center}
\end{table}

We have about a $5\%$ discrepancy between out lattice
Wilson coefficient from the two different intermediate schemes, and this will introduce about a
$10\%$ of systematic error in our final results for $\Delta M_K$. We can potentially overcome 
this by working on a finer lattice which minimize lattice artifact and using
step scaling to match at a higher scale to minimize the errors in perturbation theory.

In this calculation, we have considered the two-pion intermediate state contribution to 
$\Delta M_K$, which actually depend on the volume of the lattice. The finite volume and
infinite volume result for $\Delta M_K$ are related by \cite{Christ:finiteV}:
\begin{eqnarray}
	2\sum_{n} \frac{f ( E_{n} )}{m_{k } -E_{n}} & = & 2 \mathcal{P}  \int \tmop{dE}
	\rho_{V} ( E ) \frac{f ( E )}{m_{k } -E} + 2 \left( f ( m_{K} ) \cot ( h )
	\frac{\tmop{dh}}{\tmop{dE}} \right)_{\tmop{m_K}} \\
	f(m_K) &=& _V\langle \bar{K^0}| H_W | \pi\pi_{E=m_K}\rangle_V\,_V\langle \pi\pi_{E = m_K} | H_W | K_0\rangle _V 
\end{eqnarray}

The two-pion contribution to the $\Delta M_K$ with $I = 2$ are highly suppressed compared to 
the $I=0$ contribution, due to the $\Delta I = 1/2$ rule. We can apply \ref{Eq.ind_contr} to calculate
their individual contribution to $\Delta M_K$, with results in Table \ref{table:pipi_klks}. 
Therefore, we only consider the $I=0$ intermediate state here, with the
terms relevant for the finite volume correction in Table \ref{table:finiteV}.
\begin{table}[h]
	\centering
	\begin{tabular}{c|c|c|c} \hline
		$E_{\pi \pi I = 0} $ & $E_{\pi\pi I = 2}$ & $\Delta M_K(\pi\pi_{I = 0}) $ & 
		$\Delta M_K (\pi\pi _{I = 2})$ \\ \hline 
		336.5(15) & 346.5(9) & $-0.074(63)$ & $-6.70(7) \times 10 ^{-4} $\\\hline
	\end{tabular}
	\caption{Two-pion energy (in MeV) and their contribution to $\Delta M_K$ (in $10^{-12}$ MeV). 
	$m_c$ = 750 MeV}.
	\label{table:pipi_klks}
\end{table}

\begin{table}[h]
	\centering
	\begin{tabular}{c|c|c|c|c|c} \hline
		$\Delta M_K(\pi\pi_{I = 0})$ & 	$h = \delta + \phi$ & $\cot{h}$ &
		$dh/dE$ & $\cot{h}\times dh/dE$ & $\Delta M_K(FV)$ \\\hline
		0.084(70) & 1.71(6) & -0.144(64) & 15.3(3) & -2.2(10) & 0.019(19) \\ \hline
	\end{tabular}
	\caption{$\pi\pi_{I=0}$ contribution to $\Delta M_K$, relevant terms for finite volume correction,
	and the finite volume correction term to mass difference $\Delta M_K(FV)$, in unit of $10^{-12} MeV$.}
	\label{table:finiteV}
\end{table}
We can see that two-pion intermediate state only contribute a small fraction of the total
$\Delta M_K$, and the corresponding finite volume correction is even smaller (less than 1\%).
Therefore, being unable to measure precisely the kaon to two-pion matrix element does not give rise
to much statistical error in $\Delta M_K$.

\section{Conclusion and outlook}
We have shown that having a two-pion intermediate state to subtract does not make the 
calculation much harder, and the finite volume corrections are controllable. The results are listed 
in Table \ref{table:ind_klks}. A smaller charm mass gives rise to smaller $\Delta M_K$, because 
in all the 4-point diagrams, the charm quark enters as $u-c$. The smaller charm quark mass 
is more closer to up quark mass making the GIM cancellation more significant. 
Although $\Delta M_K$ for different $T_{min}$ agree within
errors, a noticeable difference appears if we go to $T_{min} = 8$. We expect this to be improved with better
statistics. The largest part of systematic error comes from the lattice discretization error, and
an rough estimate gives about $(m_ca)^2\approx 30 \%$. We also have a systematic error in
our Wilson Coefficient which is about $10\%$.

In future calculation, we can have more reliable results by going to physical kinematics,
with $2+1+1$ flavor finer lattice and unquenched charm quark. The RBC collaboration is now
working to generate a $80^2 \times 96 \times 192$ lattice with $1/a=3$ GeV \cite{Bob:proceeding}. 
This work was supported in part by US DOE grant
DE-SC0011941, and we thank the RIKEN BNL Research Center for the use of 
the IBM BG/Q supercomputers on which this calculation was performed.

\bibliographystyle{plain}
\bibliography{reference}

\end{document}